\documentclass[reprint,
superscriptaddress,
 amsmath,amssymb,
 aps,pra,
]{revtex4-1}
\usepackage{multirow}
\usepackage{comment}
\usepackage[caption=false,justification=justified]{subfig}
\usepackage{leftidx}
\usepackage{graphicx}
\usepackage{dcolumn}
\usepackage{bm}
\usepackage{hyperref}
\usepackage{rotating}
\usepackage{xcolor}

\newcommand{\pt}[1]{\left( #1 \right)}
\newcommand{\pq}[1]{\left[ #1 \right]}

\newcommand{\bra}[1]{\left\langle #1 \right\vert}
\newcommand{\ket}[1]{\left\vert #1 \right\rangle}

\newcommand{\Braket}[2]{\left\langle #1 \bigg\vert #2 \right\rangle}

\begin{document}
\preprint{APS/123-QED}

\title{Ultimate quantum sensitivity in the 3D relative localisation of two single-photon emitters via two-photon interference}

\author{Luca Maggio}
\affiliation{School of Mathematics and Physics, University of Portsmouth, Portsmouth PO1 3QL, United Kingdom}
\author{Vincenzo Tamma}
\email{vincenzo.tamma@port.ac.uk}
\affiliation{School of Mathematics and Physics, University of Portsmouth, Portsmouth PO1 3QL, United Kingdom}
\affiliation{Institute of Cosmology and Gravitation, University of Portsmouth, Portsmouth PO1 3FX, United Kingdom}

\date{\today}

\begin{abstract}
 We present a quantum sensing protocol for the simultaneous estimation of the difference in the localization parameters of two single-photon sources, paving the way to single-photon 3D imaging and 3D nanoscopy beyond the diffraction limit. This is achieved by exploiting two-photon interference of the two emitted photons at a beam splitter via sampling measurements in the frequency and transverse momenta at the output. We prove theoretically that this technique reaches the ultimate sensitivity in the 3D relative localization of two emitters, already with a number of sampling measurements of $\sim 1000$ and a bias in the three localization parameters below $1\%$. These results are independent of the values of the localization parameters to estimate.


\end{abstract}

\maketitle

\section{Introduction}
Two-photon interference at a beam splitter is a fundamental phenomenon at the heart of quantum optics and quantum technologies~\cite{PhysRevLett.59.2044,shih1988new}. If the two-photons fully overlap in their wave packets the probability amplitudes associated with coincidence events interfere destructively, and therefore the photons always bunch at the same detector at the beam splitter output. Therefore, by counting the coincidence events, it is possible to estimate the overlap between the two-photon wave packets, encoding specific parameters such as a temporal delay~\cite{lyons2018attosecond}, a frequency shift~\cite{PhysRevA.91.013830,Jin:15,Gianani_2018,fabre2021parameter}, or polarization~\cite{harnchaiwat2020tracking,sgobba2023optimal}. This is relevant for applications in quantum computing~\cite{kok2007linear,barz2012demonstration}, quantum key distribution~\cite{tang2014measurement,guan2015experimental}, quantum repeaters~\cite{sangouard2011quantum,hofmann2012heralded}, and quantum coherence tomography~\cite{teich2012variations}.

However, such a scheme is limited both in the precision and in the range of the values of the given parameter to estimate. More recently, such constraints were overcome by exploiting sampling measurements that resolve inner photonic parameters for the quantum estimation of temporal delays and spatial displacements~\cite{triggiani2023ultimate,triggiani2024estimation,muratore2024superresolution}.  Despite such results, the ability to estimate multiple parameters with ultimate quantum precision has remained so far challenging. To the best of the authors' knowledge, even though several 3D imaging and 3D nanoscopy protocols have been developed, the achievement of an optimal protocol, i.e. that can reach the ultimate sensitivity, has never been proven so far. Highly sensitive protocols include fusion modules~\cite{tragakis2024igaf}, adaptive optical imaging~\cite{cameron2024adaptive}, and imaging based on time-of-flight measurements~\cite{9264255}. However, these sensing schemes measure directly the localization parameters and, therefore, they are limited by the diffraction and by the precision of the detectors employed in the direct measurements.

In this Letter, we demonstrate a technique which allows one to simultaneously estimate with ultimate quantum precision the three parameters for the localization of a probe single-photon emitter with respect to a reference one. This is possible by observing the quantum interference of both photons at a beam splitter when measurements which resolve the three components of the momenta of the two photons are employed in the far field. This protocol can pave the way to new quantum technologies in 3D imaging and 3D nanoscopy~\cite{koskinen1992comparison,kolb2010time,horaud2016overview}. Applications in biological sensing can include cancer biology~\cite{chen2016imaging}, signaling~\cite{dudok2015cell}, immunology~\cite{williamson2011pre}, bacteriology~\cite{gahlmann2014exploring}, virology~\cite{hanne2016stimulated} and in medical diagnostic~\cite{benda2016sted}, since this protocol does not require exposure to a large number of photons that could damage the analyzed sample~\cite{schermelleh2019super}. 

We show that, in general, the sensitivity of the scheme decreases when one ‘classically’ averages over the information associated with the momenta of the detected photons, and it is completely lost when there is no spatial overlap between the two input photons. Only in the limit of full spatial overlap between the photons, i.e. almost equal localization parameters for both photons, it is possible to achieve the same precision by employing non-resolving measurements. On the other hand, the proposed momentum-resolved technique allows us to achieve ultimate quantum precision for any values of such parameters already with $\sim1000$ sampling measurements, independently of the value of the localization parameters.
\begin{figure}
\centering
  \includegraphics[width=80mm]{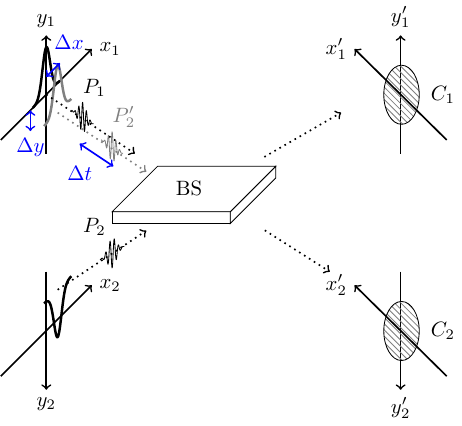} 
  \caption{3D quantum sensing scheme. Two photons, namely $P_1$ and $P_2$, are temporally delayed with respect to each other by a value $\Delta t$. Also, one is displaced with respect to the other in the transverse position so that the difference in position between $P_1$ and the symmetric image of $P_2$, namely $P'_2$, has value $\Delta x$ along the x-axis and $\Delta y$ along the y-axis. Both the photons enter in the two different input channel of a balanced beam splitter (BS). The state that represent these two photons is expressed in Eq~\ref{eq:input}. After the BS, their frequency shift and the difference in transverse momenta along the two axes is resolved by the cameras $C_1$ and $C_2$.}
  \label{fig:setup}
\end{figure}
\section{Experimental setup}
In this section we describe the experimental setup in FIG.~\ref{fig:setup}. Two independent photons impinge on a balanced beam splitter, each of them from a different input port. The two photons are prepared in such a way that their probability distributions in the momentum domain are identical, i.e. $\left\vert\psi_1(\vec{k})\right\vert^2=\left\vert\psi_2(\vec{k})\right\vert^2=\left\vert\psi(\vec{k})\right\vert^2$, where $\vec{k}=(\omega,k_x,k_y)$. Here, $k_x$ and $k_y$ refer to the transverse momenta along the x and y coordinates, $\omega$ to the frequency, and the indices 1 and 2 refer to the photon in inputs 1 and 2, respectively. From now on, we will consider the single-photon distribution in the momentum space as a Gaussian distribution that is factorable as a product of three Gaussian distributions in $k_x$, $k_y$ and $\omega$, each of them with variance $\sigma^2_{k_x}$, $\sigma^2_{k_y}$ and $\sigma^2_\omega$, respectively. However, the results discussed in this work can be generalized for every given distribution in the momentum domain.

Each photon is centered in its transverse position along the x and y coordinates in $x_j$ and $y_j$, and it is emitted at a time such that its temporal distribution is centered in $t_j$, where $j=1,2$. As a consequence, the amplitude of each photon is $\psi_j(\vec{k})=\vert\psi(\vec{k})\vert\mathrm{e}^{i\vec{k}\cdot\vec{x_j}}$, where $\vec{x_j}=(t_j,-x_j,-y_j)$. The bosonic operators in the momentum domain, i.e. $\hat{a}^\dagger_{1}\pt{\vec{k}}$ and $\hat{d}^\dagger_{2}\pt{\vec{k}}$ are defined in such a way they can take also into account the inner modes of the photons that are not related to their momentum or location. Therefore, their commutation relation is $[\hat{a}^\dagger_{1}\pt{\vec{k}_1},\hat{d}^\dagger_{2}\pt{\vec{k}_2}]=\sqrt{\nu}\delta^3(\vec{k}_1-\vec{k}_2)$, where $\nu\in[0,1]$ represents the degree of indistinguishability of the photons when momentum-resolved measurements are performed. The two-photon state thus reads (for a full description of the input state, see Appendix~\ref{app:input})
\begin{align}
\begin{split}
    \ket{\Phi}&=\int d^3k_1 \psi_1\pt{\vec{k}_1}\hat{a}^\dagger_1\pt{\vec{k}_1}\ket{0}\\ &\otimes\int d^3k_2 \psi_2\pt{\vec{k}_2}\hat{d}^\dagger_2\pt{\vec{k}_2}\ket{0}.
\end{split}\label{eq:input}
\end{align}
After impinging at the beam splitter, the two photons can be reflected or transmitted. The combination of reflection or transmission of each photon can lead to a bunching event or a coincidence event for which the photons exit the same output port or two different output ports, respectively. Then two cameras, namely $C_1$ and $C_2$ in FIG.~\ref{fig:setup}, resolve the three components of the momenta of the two photons by measuring both their frequency and transverse momenta. We measure their transverse momenta by measuring their transverse position in the far field where they are detected by either of the two cameras. This position, defined in the transverse coordinates $(x'_j,y'_j)$ (where $j$ refers to the $j$-th output port, $j=1,2$) is linked to the transverse momentum since in the far field regime, $(k_x,k_y)=\omega (x'_j,y'_j)/cd$ holds, where $d$ is the distance of the detectors from the beam splitter ($d$ is equal for both the detectors), and $c$ is the speed of light (for a mathematical description of the evolution of the probe in the sensing protocol, see Appendix~\ref{app:evolution}). 

In order to evaluate the output probabilities of having bunching, labeled with the letter B, or anti-bunching, labeled with the letter A, it is useful to introduce the following parameters
\begin{align}
\begin{split}
    \xi=\frac{\Delta\omega}{2\sigma_\omega}&, \qquad\tau=2\sigma_\omega\Delta t,\\
    \kappa_x=\frac{\Delta k_x}{2\sigma_{k_x}}&, \qquad\lambda_x=2\sigma_{k_x}\Delta x,\\
    \kappa_y=\frac{\Delta k_y}{2\sigma_{k_y}}&,\qquad\lambda_y=2\sigma_{k_y}\Delta y,\label{eq:notation}
    \end{split}
\end{align}

where $\Delta\omega=\omega_2-\omega_1$, $\Delta k_x=k_{x,2}-k_{x,1}$, $\Delta k_y=k_{y,2}-k_{y,1}$, $\Delta t=t_2-t_1$, $\Delta y=y_2-y_1$ and $\Delta x=x_2-x_1$. Such probabilities thus read (for detailed derivations, see Appendices~\ref{app:coincidence}, \ref{app:bunching} and \ref{app:prob})

\begin{align}
    \begin{split}
          P_\nu\pt{X;\vec{\kappa}\vert \vec{\lambda}}=\frac{\gamma^2}{(\pi)^{3/2}}\mathrm{e}^{-\vert\vec{\kappa}\vert^2}\zeta_{X;\nu}(s\vec{\kappa}\cdot\vec{\lambda}).\label{eq:prob}
    \end{split}
\end{align}
Here, $\zeta_{X;\nu}(s\vec{\kappa}\cdot\vec{\lambda})$ represents the quantum beating,
\begin{equation}
    \zeta_{X;\nu}(s\vec{\kappa}\cdot\vec{\lambda})=\frac{1}{2}\left(1+\alpha\pt{X}\nu\cos\pt{s\vec{\kappa}\cdot\vec{\lambda}}\right),\label{eq:quantumbeats}
\end{equation}
where $\vec{\kappa}=(\xi,\kappa_x,\kappa_y)$, 
   \begin{equation}
   \vec{\lambda}= \frac{1}{\sqrt{\tau^2+\lambda_x^2+\lambda_y^2}}\begin{pmatrix}
        \tau\\-\lambda_x\\-\lambda_y
    \end{pmatrix}=\begin{pmatrix}
        \cos(\theta)\\\sin(\theta)\cos(\phi)\\\sin(\theta\sin(\phi))
    \end{pmatrix},\label{eq:unitaryvec}
\end{equation}

$s=\sqrt{\tau^2+\kappa_x^2+\kappa_y^2}$, $X=A,B$, $\alpha(A)=-1$, $\alpha(B)=1$ and $\gamma\in[0,1]$ is the detector efficiency. 

This sensing protocol can estimate the independent parameters $s,\theta,\phi$ simultaneously. The precision in measuring the transverse momenta in the far field, therefore the precision in the output transverse position, namely $(\delta x',\delta y')$, as well as the frequency precision $\delta\omega$, must be small enough to resolve the distribution in the momentum space and the quantum beatings, namely
\begin{equation}
        \delta \vec{\kappa}\ll 1,\qquad\delta \vec{\kappa}\cdot\vec{\lambda}\ll \frac{1}{s}.\label{eq:conditions}
\end{equation}

Such conditions are independent of the wavelength of the probes. Therefore, the estimation is not affected by the diffraction limit~\cite{rayleigh1879xxxi,born2013principles}. 

If the experimental setup does not resolve in the momentum space, we can retrieve only one parameter. In this case, the output of the sensing scheme can be defined only as a bunching event or a coincidence event. We can find the probability of the outputs by integrating Eq.~\ref{eq:prob} with respect of the three components of $\vec{\kappa}$, obtaining
\begin{align}
\begin{split}
     \mathcal{P}_\nu\pt{X\vert s}=\frac{\gamma^2}{2}\left(1+\alpha(X)\nu\exp\left[-s^2/4\right]\right).\label{eq:non-resprob}
 \end{split}
\end{align}
Therefore, with the non-resolving scheme, we can estimate only $s$.   
\begin{figure}
\centering
 \includegraphics[width=80mm]{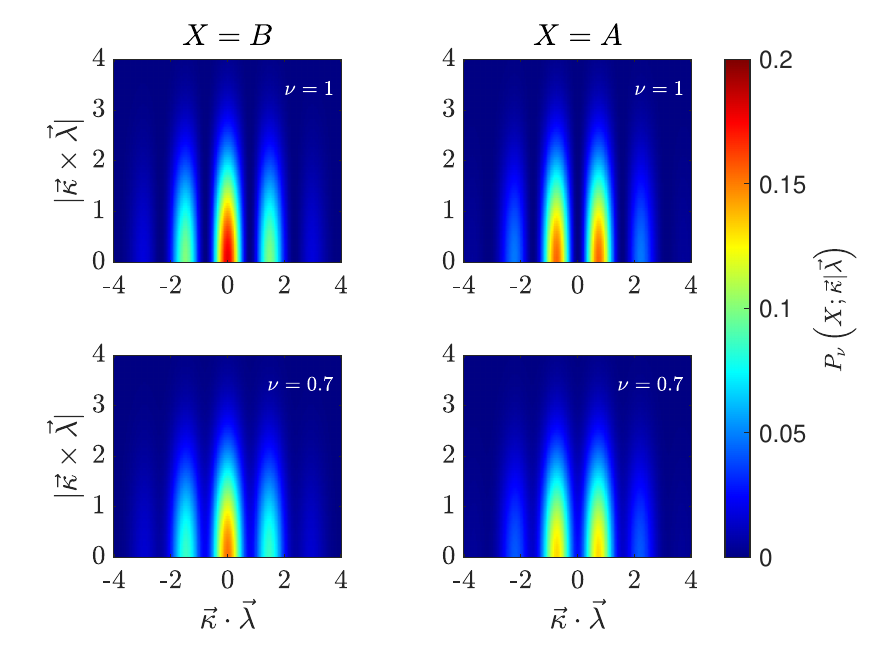}
 \caption{Plot of the probability distribution in Eq.~\ref{eq:prob} in the momentum space for photons with Gaussian momentum distribution $\vert\psi(\vec{k})\vert^2$ with vector $\vec{k}=(\omega,k_x,k_y)$. The probability has quantum beats which period depends on $\vec{\kappa}\cdot\vec{\lambda}$ and a Gaussian envelope that depends on $\vert\vec{\kappa}\vert^2=\vert\vec{\kappa}\cdot\vec{\lambda}\vert^2+\vert\vec{\kappa}\times\vec{\lambda}\vert^2$. For simplicity, we impose $\gamma=1$, $s=4$.}
\label{fig:probs}
\end{figure}

\section{Ultimate 3D quantum sensitivity}
A set of unbiased estimators $(\tilde{s}, \tilde{\theta},\tilde{\phi})$ of the localization parameters is affected by an error described by the covariance matrix $\mathrm{Cov}[\tilde{s}, \tilde{\theta},\tilde{\phi}]$. The covariance matrix is bounded below by the Cram\'{e}r-Rao bound, which is inversely proportional to the number of sampling measurements $N$, and directly proportional to the inverse of the Fisher information matrix $F_\nu(s,\theta,\phi)$~\cite{cramer1999mathematical, rohatgi2015introduction}. This bound represents the maximum precision achievable by the sensing scheme. The inequality between the covariance matrix and the  Cram\'{e}r-Rao bound means that the difference between these matrices is positive semi-definite. Ultimately, the Cram\'{e}r-Rao bound is bounded below by the quantum Cram\'{e}r-Rao bound, which is inversely proportional to the number of sampling measurements $N$, and directly proportional to the inverse of the quantum Fisher information matrix $Q(s,\theta,\phi)$~\cite{helstrom1969quantum,holevo2011probabilistic}. Instead, this bound represents the maximum precision achievable in nature for the estimation of the parameters $(s,\theta,\phi)$. Similarly as before, this inequality means that the difference between Cram\'{e}r-Rao bound and quantum Cram\'{e}r-Rao bound is positive semi-definite. To summarize, this set of inequalities hold
\begin{equation}
\mathrm{Cov}[\tilde{s}, \tilde{\theta},\tilde{\phi}]\geq \frac{F^{-1}_\nu\pt{s,\theta,\phi}}{N}\geq \frac{Q^{-1}\pt{s,\theta,\phi}}{N}, \label{eq:bounds}
\end{equation}
where, given the state $\ket{\Phi}$ in Eq.~\ref{eq:input}, the quantum Fisher information matrix is (for a detailed derivation, see Appendix~\ref{app:QFI})
\begin{equation}
    Q(s,\theta,\phi)=\frac{1}{2}\begin{pmatrix}
        1&0&0\\0&s^2&0\\0&0&s^2\sin^2\theta
    \end{pmatrix}.\label{eq:Q}
\end{equation}

The Fisher information matrix can be obtained from the probabilities in Eq.~\ref{eq:prob}. In particular, by considering $\nu=1$, we obtain
\begin{align}
F_{\nu=1}\pt{s,\theta,\phi}=\gamma^2Q\pt{s,\theta,\phi}\label{eq:finu1}
\end{align}
This means that in the limit of perfect detectors and identical photons in all degrees of freedom but the localization parameters, it is possible to achieve the ultimate precision for the estimation of $(s,\theta,\phi)$. Since the Fisher information matrix is diagonal, we notice that in the asymptotic limit of large $N$ there is no correlation between the estimation of the parameters. When we do not resolve in the momentum space, we can obtain indistinguishability only if $s\sim 0$. Under this assumption and while having perfect detectors, the Fisher information for the non-resoling protocol is 
\begin{equation}
    \mathcal{F}_{\nu=1}(s)\xrightarrow{s\rightarrow 0}\gamma^2Q_{ss}(s,\theta,\phi)=\frac{\gamma^2}{2}.\label{eq:finu1s0}
\end{equation}
Therefore, in this limit $s\rightarrow 0$ it is possible to achieve ultimate precision in the estimation of only the parameter $s$ without resolving the photonic momenta at the detection.

\section{3D localization in the case of photons partially distinguishable at the detectors}

\begin{figure}
\centering
\includegraphics[width=80mm]{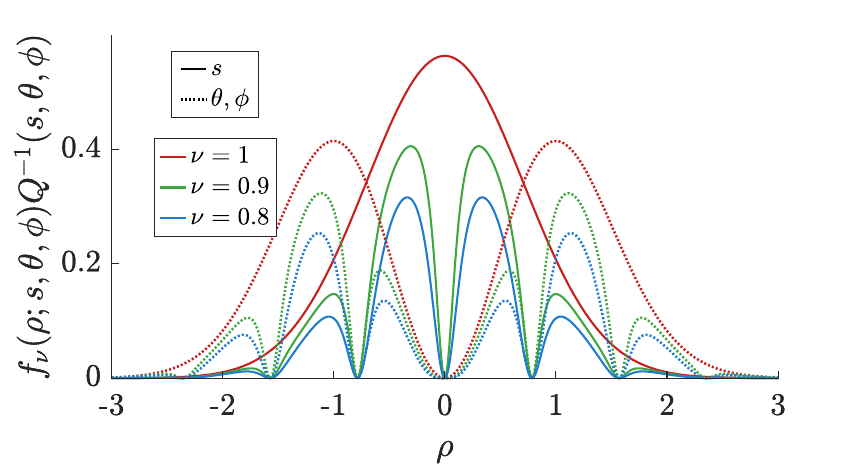}
\caption{Contributions of the Fisher information matrix density $f_{\nu}\pt{\rho;s,\theta,\phi}$ in Eq~\ref{eq:densityfi} as a function of $\rho=\vec{\lambda}\cdot\vec{\kappa}$ for different values of the visibility parameter $\nu=1,0.9,0.8$ for photons with Gaussian momentum distribution $\vert\psi(\vec{k})\vert^2$ with vector $\vec{k}=(\omega,k_x,k_y)$. For simplicity, we impose $s=4$.}
\label{fig:cont}
\end{figure}
In the case of photons that differ in other parameters apart from the localization ones, that is, when $\nu\neq 1$, this sensing protocol is still able to estimate the parameters $(s,\theta,\phi)$ although without saturating the quantum Cram\'{e}r-Rao bound. Indeed, in this scenario, the quantum beats lose visibility, as shown in  FIG.~\ref{fig:probs}. The Fisher information matrix is an integral with respect to the resolved parameters that defines the vector $\vec{\kappa}$. In particular, if we rename the parameter of integration $\vec{\lambda}\cdot\vec{\kappa}=\rho$, we have 
\begin{align}
F_{\nu}(s,\theta,\phi)&=\gamma^2\int d\rho f_\nu(\rho;s,\theta,\phi),\label{eq:Fi}
\end{align}
where $f_\nu(\rho;s,\theta,\phi)$ is the contribution for each value $(s,\theta,\phi)$ to the Fisher information as plotted in FIG.\ref{fig:cont} (for the evaluation of the Fisher information matrix as a function of the distinguishability parameter $\nu$, see Appendix~\ref{app:FI}). It reads
\begin{equation}
    f_\nu(\rho;s,\theta,\phi)= \frac{\mathrm{e}^{-\rho^2}\beta_\nu(s\rho)}{(\pi)^{1/2}}\begin{pmatrix}
        \rho^2&0&0\\
        0&\frac{s^2}{2}&0\\
        0&0&\frac{s^2\sin^2\theta}{2}
    \end{pmatrix}\label{eq:densityfi},
\end{equation}

where the function
\begin{align}
\begin{split}
    \beta_\nu(s\rho)&=\sum_{X=A,B}\frac{1}{\zeta_{X;\nu}(s\rho)}\left(\frac{\partial\zeta_{X;\nu}(x)}{\partial x}\right)_{x=s\rho}^2\\&=\frac{\nu^2\sin^2(s\rho)}{1-\nu^2\cos^2(s\rho)}
    \end{split}\label{eq:beta}
\end{align}
describes the contributions to the fisher information from the quantum beats emerging in the expression of $\zeta_{X;\nu}(s\rho)$ in Eq.~\ref{eq:quantumbeats}.

If $\beta_\nu(s\rho)=0$,  no parameters can be estimated (which happens for $\nu=0$, and therefore when the two photons are fully distinguishable at the detectors). Instead, for $\beta_\nu(s\rho)=1$, the precision in the estimation is maximum (that is, when $\nu=1$, meaning that all the degrees of freedom of the photons but the ones to estimate are identical). These properties of $\beta_\nu(s\rho)$ are linked to the fact that $\beta_\nu(s\rho)$ is a function of $(\partial \zeta_{X;\nu}(x)/\partial x)_{x=s\rho}$, for $X=A,B$, which represents the degree of sensitivity at small variation of the value $s\rho$ defined by the localization parameters to estimate.

Another important property of $\beta_\nu(s\rho)$ is that it oscillates with half the period and in the same direction as the beatings of the probabilities in Eq.~\ref{eq:prob} as emerging from the dependence on  the function $(\partial \zeta_{X;\nu}(x)/\partial x)_{x=s\rho}^2=\nu^2\sin^2(s\rho)/4=\nu^2(1-\cos(2s\rho))/8$ which describes the momentum-dependent sensitivity of our protocol.  As expected, the sensitivity is lost for $\nu\neq1$ at the stationary points of $\zeta_{X;\nu}(s\rho)$ for values of $s\rho$ multiple of $\pi$.

\begin{figure*}
\begin{minipage}[c]{0.48\linewidth}
\includegraphics[width=\linewidth]{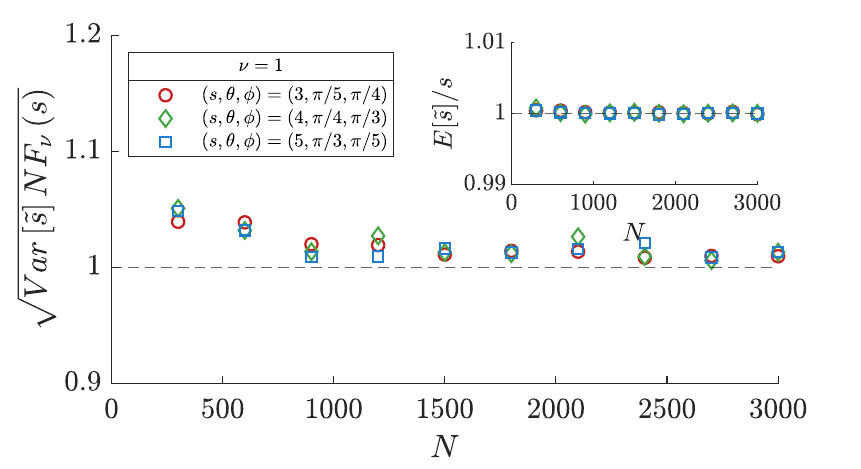}
\includegraphics[width=\linewidth]{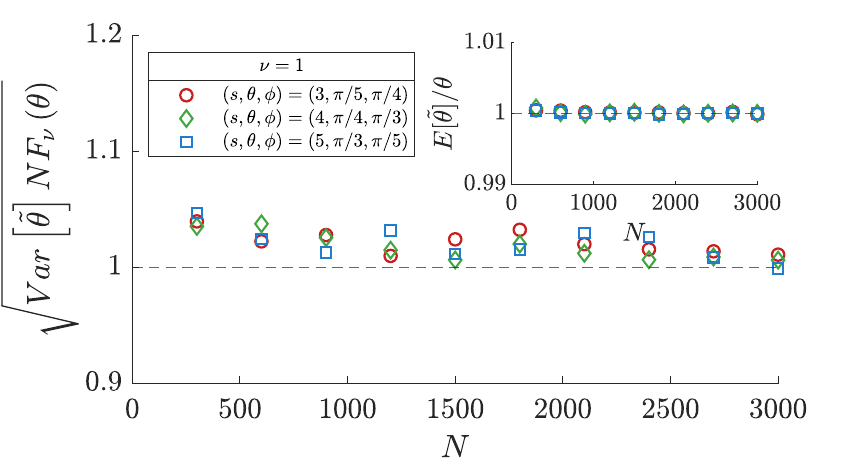}
\includegraphics[width=\linewidth]{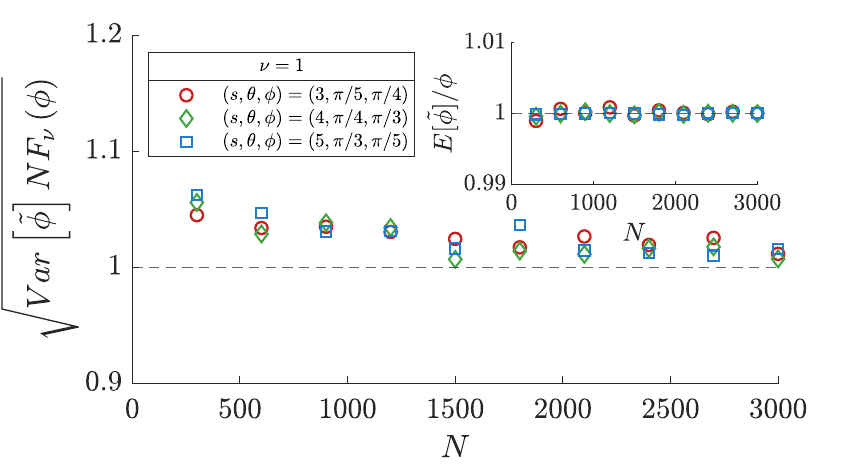}

\end{minipage}
\hfill
\begin{minipage}[c]{0.48\linewidth}
\includegraphics[width=\linewidth]{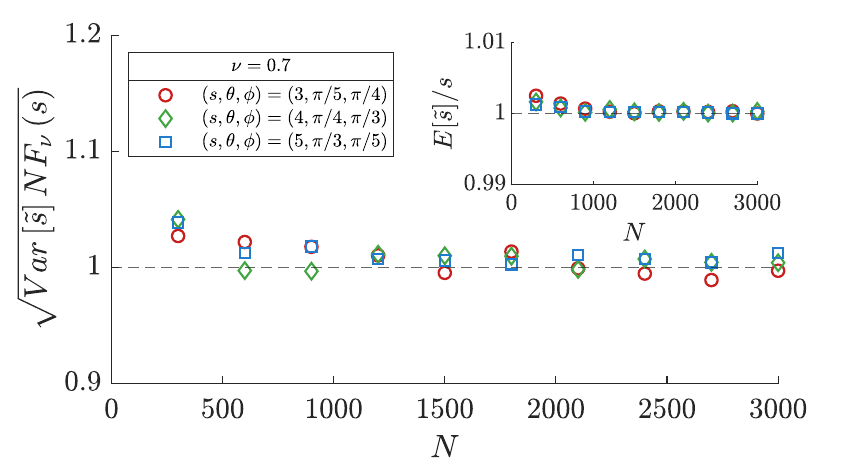}
\includegraphics[width=\linewidth]{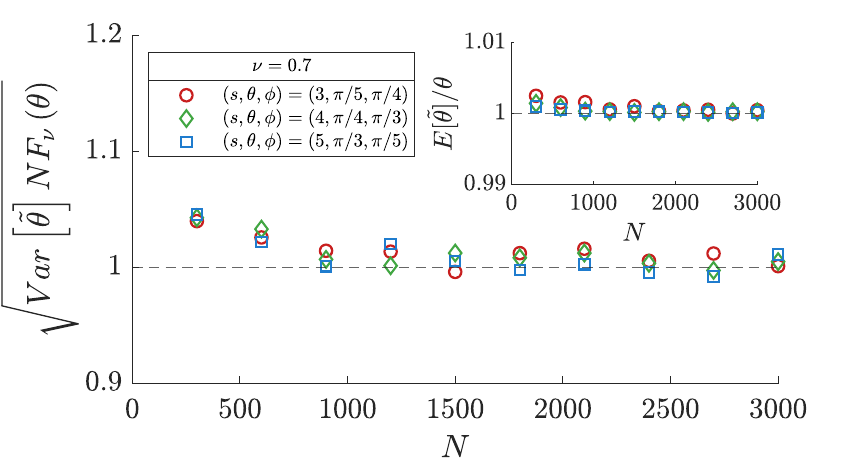}
\includegraphics[width=\linewidth]{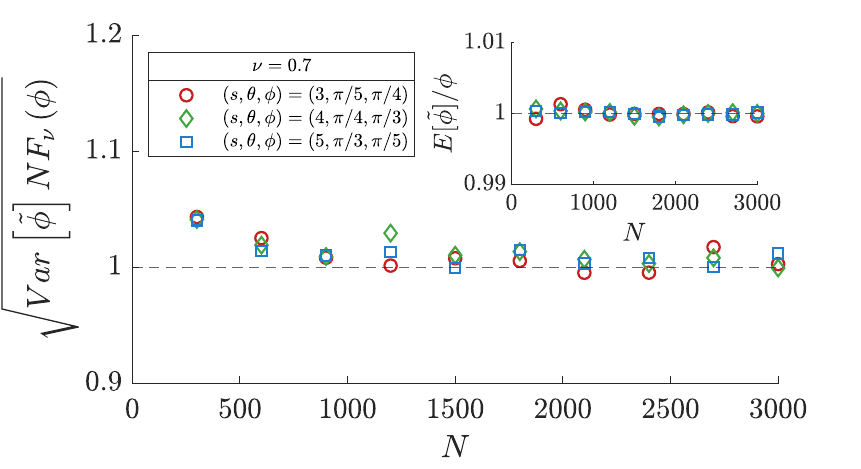}

\end{minipage}%
\caption{Simulations of the variance for the estimation of the parameters $s,\theta,\phi$ normalized with respect to the Cram\'{e}r-Rao bound in function of the number of repetition of the measurement $N$. For this simulations, we consider two values of the visibility $\nu=1,0.7$ and the three following circumstances: $(s,\theta,\phi)=(3,\pi/5,\pi/4),(4,\pi/4,\pi/3),(5,\pi/3,\pi/5)$. In the insets we show the simulations of the expected value of the estimators $(\tilde{s},\tilde{\theta},\tilde{\phi})$ normalized. We execute the simulations by repeating for each point the measure process $10^4$ times. It is possible to show that even for a number of sampling measurements $N\sim 10^3$ the Cram\'{e}r-Rao bound is approximately saturated, and the bias is inferior to the $1\%$.}\label{fig:sim}
\end{figure*}

We simulate the sensitivity of this protocol in FIG~\ref{fig:sim}, where we show the variance normalized to the Cram\'{e}r-Rao bound as a function of the size of the sample $N$. Each point in this figure is obtained by finding the likelihood estimators of the parameters $(s,\theta,\phi)$ (for a more detailed discussion about the likelihood estimators, see Appendix~\ref{app:Likelihood})~\cite{rohatgi2015introduction}. In particular, each point represents an average over $1000$ attempted iterations of the estimation. From the plots it is possible to conclude that the Cram\'{e}r-Rao bound is approximately satured for a number of observed samples $N\approx 1000$. In addition, in the insets, we show the average of the values of the localization parameters in the maximum likelihood estimation normalized to the real values as a function of $N$. From the insets, we can see that the estimators are unbiased, independently of the values of the parameters to estimate, i.e. $(s,\theta,\phi)$, and of the visibility parameter $\nu$.

\section{Conclusion}
In this work, we present a 3D quantum sensing technique which allows us to simultaneously estimate with ultimate quantum precision the three parameters for the localization of a probe single-photon emitter with respect to a reference one, paving the way to single-photon 3D imaging and 3D nanoscopy beyond the diffraction limit. We show that the ultimate precision can be achieved by resolving the parameters in the 3D momentum space. Instead, by employing non-resolved measurements, only a single parameter can be estimated, and the maximum precision for its estimation is achieved under a specific condition, that is, for almost equal localization parameters for both photons.


The efficiency of the presented protocol saturates the quantum Cram\'{e}r-Rao bound, by employing non-entangled photons differing only in the localization parameters. To the best of the authors' knowledge, this is the first time that the achievement of the ultimate precision limit has been proven for a 3D imaging protocol. We also demonstrate that 3D localization can also be achieved in the case of partial distinguishability of the photons at the detectors in parameters other than the ones we wish to estimate.

Remarkably, unbiased estimation can be achieved with any number $N$ of sampling measurements and the Cram\'{e}r-Rao bound in sensitivity is saturated already for $N\sim 1000$ sampling measurements for different values of photonic distinguishability at the detectors. Therefore, this work can pave the way to 3D quantum imaging technology at the nanoscale for engineering and biomedical applications.

\acknowledgments
This project is partially supported by Xairos Systems Inc. VT acknowledges support from the Air Force office of Scientific Research under award number FA8655-23-1-7046.

\bibliography{main}

\newpage
\onecolumngrid
\appendix
\section{Input state in Eq.\ref{eq:input}}
\label{app:input}
We consider two photons that impinge on a balanced beam splitter from its two input ports. For simplicity, we describe these two photons in their transverse propagation dimension. The j-th photon is in the j-th input channel, it has a amplitude $\psi_j\pt{\vec{k}}$, where $\vec{k}=(\omega,k_x,k_y)$, defined by the frequency $\omega$ and the transverse momentum $(k_x,k_y)$, it is centered in $(x_j,y_j)$ and it impinge on the j-th input port of the beam splitter at a time $t_j$ ($j=1,2$), so that we can define the vector of the localization parameters $\vec{x}=(t_j, -x_j,-y_j)$, such that
\begin{equation}
    \psi_j\pt{\vec{k}}=\mathrm{e}^{i \vec{x}_j\cdot\vec{k}}\left\vert\psi\pt{\vec{k}}\right\vert.
\end{equation}
For a given channel j, a transverse momentum $(k_{x,j},k_{y,j})$ and a frequency $\omega_j$, the creation operators $\hat{a}^\dagger_j\pt{\vec{k}_j}$ and $\hat{b}^\dagger_j\pt{\vec{k}_j}$ denote two orthogonal modes ($j=1,2$). They are defined by the commutation relations
\begin{align}
    \begin{split}
&\pq{\hat{a}_i\pt{\vec{k}_i},\hat{a}^\dagger_j\pt{\vec{k}_j}}=\delta_{i,j}\delta^3\pt{\vec{k}_i-\vec{k}_j},\\
&\pq{\hat{b}_i\pt{\vec{k}_i},\hat{b}^\dagger_j\pt{\vec{k}_j}}=\delta_{i,j}\delta^3\pt{\vec{k}_i-\vec{k}_j},\\
&\pq{\hat{b}_i\pt{\vec{k}_i},\hat{a}^\dagger_j\pt{\vec{k}_j}}=0.
    \end{split}
\end{align}
We introduce a new creation operator
\begin{equation}
\hat{d}^\dagger_{i}\pt{\vec{k}_i}=\sqrt{\nu}\hat{a}^\dagger_{i}\pt{\vec{k}_i}+\sqrt{1-\nu}\hat{b}^\dagger_{i}\pt{\vec{k}_i}, 
\end{equation}
such that $\nu\in\pq{0,1}$. For $\nu=1$, we have $\hat{d}^\dagger_{i}\pt{\vec{k}_i}=\hat{a}^\dagger_{i}\pt{\vec{k}_i}$, while for $\nu=0$, the two operators are orthogonal. The parameter $\nu$ describes the difference of the photons in all the degrees of freedom but the one to measure, e.g. the polarization. Therefore, the input state is
\begin{align}
    \begin{split}
        \ket{\Phi}=\int d^3k_1 \psi_1\pt{\vec{k}_1}\hat{a}^\dagger_1\pt{\vec{k}_1}\ket{0}\otimes\int d^3k_2 \psi_2\pt{\vec{k}_2}\hat{d}^\dagger_2\pt{\vec{k}_2}\ket{0},\label{app:inputstate}
    \end{split}
\end{align}
which is Eq.~\ref{eq:input} in the main text, where $d^3k_i=d\omega_idk_{x,i}dk_{y,i}$, $i=1,2$.

\section{Evaluation of Eqs.\ref{eq:prob},\ref{eq:quantumbeats},\ref{eq:conditions},\ref{eq:non-resprob}}
In this section we evaluate the output probabilities of the 3D sensing protocol and the output probabilities of the non-resolving sensing scheme. In order to do so, we first describe the evolution of the two-photon input state in Eq.~\ref{app:inputstate} (see Appendix~\ref{app:evolution}). Then we find the probability of having coincidence and bunching (see Appendices~\ref{app:coincidence} and~\ref{app:bunching}, respectively). Finally, we write the general formula of the output probabilities, even for the non-resolving scheme (see Appendix~\ref{app:prob}).
\subsection{State evolution}
\label{app:evolution}
The two input photons impinge on a balance beam splitter, whose effect on the state $\ket{\Phi}$ is described by the operator $\hat{U}_{BS}$. This operator acts on the injected probe through the map $\hat{U}_{BS}\hat{a}^\dagger_i\hat{U}_{BS}=\sum_j U_{BS,ij}\hat{a}^\dagger_j$, and $\hat{U}_{BS}\hat{b}^\dagger_i\hat{U}_{BS}=\sum_j U_{BS,ij}\hat{b}^\dagger_j$, where
\begin{equation}
    U_{BS}=\frac{1}{\sqrt{2}}\begin{pmatrix}
        1&-1\\1&1
    \end{pmatrix}.
\end{equation}
Therefore, the two-photon state after the beam splitter is
\begin{align}
    \hat{U}_{BS}\ket{\Phi}=\int d^3k_1 \psi_1\pt{\vec{k}_1}\frac{\hat{a}^\dagger_1\pt{\vec{k}_1}-\hat{a}^\dagger_2\pt{\vec{k}_1}}{\sqrt{2}}\ket{0}\otimes\int d^3k_2 \psi_2\pt{\vec{k}_2}\frac{\hat{d}^\dagger_1\pt{\vec{k}_2}+\hat{d}^\dagger_2\pt{\vec{k}_2}}{\sqrt{2}}\ket{0}.
\end{align}
This state is defined by three contributes. In fact we can write
\begin{align}
    \begin{split}
        \hat{U}_{BS}\ket{\Phi}&=\frac{1}{2}\int d^3k_1 d^3k_2\psi_1\pt{\vec{k}_1}\psi_2\pt{\vec{k}_2}\hat{a}^\dagger_1\pt{\vec{k}_1}\hat{d}^\dagger_1\pt{\vec{k}_2}\ket{0}+\\
        &-\frac{1}{2}\int d^3k_1 d^3k_2\psi_1\pt{\vec{k}_1}\psi_2\pt{\vec{k}_2}\hat{a}^\dagger_2\pt{\vec{k}_1}\hat{d}^\dagger_2\pt{\vec{k}_2}\ket{0}+\\
        &+\frac{1}{2}\int d^3k_1 d^3k_2\psi_1\pt{\vec{k}_1}\psi_2\pt{\vec{k}_2}\pt{\hat{a}^\dagger_1\pt{\vec{k}_1}\hat{d}^\dagger_2\pt{\vec{k}_2}-\hat{a}^\dagger_2\pt{\vec{k}_1}\hat{d}^\dagger_1\pt{\vec{k}_2}}\ket{0},
    \end{split}
\end{align}
where the first and second rows represent, respectively, a bunching (B) on the first and second channel, and the third row represents an anti-bunching (A). After the beam splitter, the photons reaches the detectors, that are at a fixed distance $d$ from the beam splitter. We impose that the transverse momentum of each photon is much smaller than its longitudinal momentum.  In the far field regime, i.e. when the photons reach the detectors, the amplitude of each photon becomes $\psi_j^{FF}\pt{\vec{k}}$ by taking an extra phase. We obtain
\begin{equation}
    \psi_j^{FF}\pt{\vec{k}}=\mathrm{e}^{i\omega d/c}\psi_j\pt{\vec{k}}=\mathrm{e}^{i\omega d/c+i\vec{x}_j\cdot\vec{k}}\left\vert \psi\pt{\vec{k}} \right\vert,
\end{equation}
where the transverse momentum $(k_x,k_y)$ is linked to the position on the detection plane $(x'_j,y'_j)$ by the relation $(k_x,k_y)=(x'_j,y'_j)\omega/cd$,
and the two-photon state in proximity of the detectors is:
\begin{align}
    \begin{split}
        \ket{\Phi^{FF}}&=\frac{1}{2}\int d^3k_1 d^3k_2\psi_1^{FF}\pt{\vec{k}_1}\psi_2^{FF}\pt{\vec{k}_2}\hat{a}^\dagger_1\pt{\vec{k}_1}\hat{d}^\dagger_1\pt{\vec{k}_2}\ket{0}+\\
        &-\frac{1}{2}\int d^3k_1 d^3k_2\psi_1^{FF}\pt{\vec{k}_1}\psi_2^{FF}\pt{\vec{k}_2}\hat{a}^\dagger_2\pt{\vec{k}_1}\hat{d}^\dagger_2\pt{\vec{k}_2}\ket{0}+\\
        &+\frac{1}{2}\int d^3k_1 d^3k_2\psi_1^{FF}\pt{\vec{k}_1}\psi_2^{FF}\pt{\vec{k}_2}\pt{\hat{a}^\dagger_1\pt{\vec{k}_1}\hat{d}^\dagger_2\pt{\vec{k}_2}-\hat{a}^\dagger_2\pt{\vec{k}_1}\hat{d}^\dagger_1\pt{\vec{k}_2}}\ket{0}.
    \end{split}
\end{align}
\subsection{Coincidence-event probability}
\label{app:coincidence}
The detectors are sensitive to the frequency and the position of each photon. For each couple of data-frequency and position- extracted from the i-th photon, only one transverse momentum is possible. For the evaluation of the output probabilities, we first consider no losses in detection and perfect coupling efficiency between the elements of the setup. The probability of having an anti-bunching event, while measuring $\pt{\vec{k}_1,\vec{k}_2}$, is 
\begin{align}
    \begin{split}
        P\pt{A;\vec{k}_1,\vec{k_2}\vert \vec{x}_1,\vec{x_2}}&=\left\vert \bra{0}\hat{a}_1\pt{\vec{k}_1}\hat{a}_2\pt{\vec{k}_2}\ket{\Phi^{FF}}\right\vert^2+\\
        &+\left\vert \bra{0}\hat{a}_1\pt{\vec{k}_1}\hat{b}_2\pt{\vec{k}_2}\ket{\Phi^{FF}}\right\vert^2+\\&+\left\vert \bra{0}\hat{b}_1\pt{\vec{k}_1}\hat{a}_2\pt{\vec{k}_2}\ket{\Phi^{FF}}\right\vert^2+\\&+\left\vert \bra{0}\hat{b}_1\pt{\vec{k}_1}\hat{b}_2\pt{\vec{k}_2}\ket{\Phi^{FF}}\right\vert^2.
    \end{split}
\end{align}
We evaluate each term of this probability as follows:
\begin{align}
    \begin{split}
      \left\vert \bra{0}\hat{a}_1\pt{\vec{k}_1}\hat{a}_2\pt{\vec{k}_2}\ket{\Phi^{FF}}\right\vert^2&=\left\vert\frac{\sqrt{\nu}}{2}\pt{\psi_2^{FF}\pt{\vec{k}_2}\psi_1^{FF}\pt{\vec{k}_1}-\psi_2^{FF}\pt{\vec{k}_1}\psi_1^{FF}\pt{\vec{k}_2}} \right\vert^2 \\
      &=\frac{\nu}{4}\left\vert \psi\pt{\vec{k}_1} \psi\pt{\vec{k}_2} \right\vert^2\left\vert \mathrm{e}^{i\vec{x}_1\cdot\vec{k}_1}\mathrm{e}^{i\vec{x}_2\cdot\vec{k}_2}-\mathrm{e}^{i\vec{x}_1\cdot\vec{k}_2}\mathrm{e}^{i\vec{x}_2\cdot\vec{k}_1}\right\vert^2\\
      &=\frac{\nu}{2}\left\vert \psi\pt{\vec{k}_1} \psi\pt{\vec{k}_2} \right\vert^2\pq{1-\cos\pt{\Delta\vec{x}\cdot\Delta\vec{k}}},
    \end{split}
\end{align}
\begin{align}
    \begin{split}
        \left\vert \bra{0}\hat{a}_1\pt{\vec{k}_1}\hat{b}_2\pt{\vec{k}_2}\ket{\Phi^{FF}}\right\vert^2&=\left\vert\frac{\sqrt{1-\nu}}{2}\psi_2^{FF}\pt{\vec{k}_2}\psi_1^{FF}\pt{\vec{k}_1}\right\vert^2\\
        &=\frac{1-\nu}{4}\left\vert \psi\pt{\vec{k}_1} \psi\pt{\vec{k}_2}\right\vert^2,
    \end{split}
\end{align}
\begin{align}
    \begin{split}
        \left\vert \bra{0}\hat{b}_1\pt{\vec{k}_1}\hat{a}_2\pt{\vec{k}_2}\ket{\Phi^{FF}}\right\vert^2&=\left\vert\frac{\sqrt{1-\nu}}{2}\psi_1^{FF}\pt{\vec{k}_2}\psi_2^{FF}\pt{\vec{k}_1}\right\vert^2\\
        &=\frac{1-\nu}{4}\left\vert \psi\pt{\vec{k}_1} \psi\pt{\vec{k}_2}\right\vert^2,
    \end{split}
\end{align}
\begin{equation}
    \left\vert \bra{0}\hat{b}_1\pt{\vec{k}_1}\hat{b}_2\pt{\vec{k}_2}\ket{\Phi^{FF}}\right\vert^2=0.
\end{equation}
The sum of all the contributions is
\begin{equation}
     P\pt{A;\vec{k}_1,\vec{k_2}\vert \vec{x}_1,\vec{x_2}}=\frac{1}{2}\left\vert \psi\pt{\vec{k}_1} \psi\pt{\vec{k}_2} \right\vert^2\pq{1-\nu\cos\pt{\Delta\vec{x}\cdot\Delta\vec{k}}}.
\end{equation}

\subsection{Bunching-event probability}
\label{app:bunching}
The probability of having a bunching event, while measuring $\pt{\vec{k}_1,\vec{k_2}}$, is
\begin{align}
    \begin{split}
        P\pt{B;\vec{k}_1,\vec{k_2}\vert \vec{x}_1,\vec{x_2}}&=\frac{1}{2}\sum_{i=1,2}\left\vert \bra{0}\hat{a}_i\pt{\vec{k}_1}\hat{a}_i\pt{\vec{k}_2}\ket{\Phi^{FF}}\right\vert^2+\\
        &+\sum_{i=1,2}\left\vert \bra{0}\hat{a}_i\pt{\vec{k}_1}\hat{b}_i\pt{\vec{k}_2}\ket{\Phi^{FF}}\right\vert^2+\\&+\frac{1}{2}\sum_{i=1,2}\left\vert \bra{0}\hat{b}_i\pt{\vec{k}_1}\hat{b}_i\pt{\vec{k}_2}\ket{\Phi^{FF}}\right\vert^2
    \end{split}
\end{align}
In analogy to the previous subsection, we evaluate each term involved in the evaluation of the probability, obtaining
\begin{align}
    \begin{split}
        \frac{1}{2}\sum_{i=1,2}\left\vert \bra{0}\hat{a}_i\pt{\vec{k}_1}\hat{a}_i\pt{\vec{k}_2}\ket{\Phi^{FF}}\right\vert^2&=\left\vert\frac{\sqrt{\nu}}{2}\pt{\psi_2^{FF}\pt{\vec{k}_2}\psi_1^{FF}\pt{\vec{k}_1}+\psi_2^{FF}\pt{\vec{k}_1}\psi_1^{FF}\pt{\vec{k}_2}} \right\vert^2 \\
      &=\frac{\nu}{4}\left\vert \psi\pt{\vec{k}_1} \psi\pt{\vec{k}_2} \right\vert^2\left\vert \mathrm{e}^{i\vec{x}_1\cdot\vec{k}_1}\mathrm{e}^{i\vec{x}_2\cdot\vec{k}_2}+\mathrm{e}^{i\vec{x}_1\cdot\vec{k}_2}\mathrm{e}^{i\vec{x}_2\cdot\vec{k}_1}\right\vert^2\\
      &=\frac{\nu}{2}\left\vert \psi\pt{\vec{k}_1} \psi\pt{\vec{k}_2} \right\vert^2\pq{1+\cos\pt{\Delta\vec{x}\cdot\Delta\vec{k}}},
    \end{split}
\end{align}
\begin{align}
\begin{split}
    \sum_{i=1,2}\left\vert \bra{0}\hat{a}_i\pt{\vec{k}_1}\hat{b}_i\pt{\vec{k}_2}\ket{\Phi^{FF}}\right\vert^2&=\sum_{i=1,2}\left\vert\frac{\sqrt{1-\nu}}{2}\psi_i^{FF}\pt{\vec{k}_2}\psi_i^{FF}\pt{\vec{k}_1}\right\vert^2\\
        &=\frac{1-\nu}{2}\left\vert \psi\pt{\vec{k}_1} \psi\pt{\vec{k}_2}\right\vert^2,
        \end{split}
\end{align}
\begin{align}
    \begin{split}
        \frac{1}{2}\sum_{i=1,2}\left\vert \bra{0}\hat{b}_i\pt{\vec{k}_1}\hat{b}_i\pt{\vec{k}_2}\ket{\Phi^{FF}}\right\vert^2=0.
    \end{split}
\end{align}
Therefore, the bunching probability is
\begin{equation}
    P\pt{B;\vec{k}_1,\vec{k_2}\vert \vec{x}_1,\vec{x_2}}=\frac{1}{2}\left\vert \psi\pt{\vec{k}_1} \psi\pt{\vec{k}_2} \right\vert^2\pq{1+\nu\cos\pt{\Delta\vec{x}\cdot\Delta\vec{k}}}.
\end{equation}
\subsection{Evaluation of the output probabilities under the condition of non-unitary efficiency detection}
\label{app:prob}
To summarize, the probability of having bunching or anti-bunching while measuring $\pt{\vec{k}_1,\vec{k}_2}$ is
\begin{equation}
    P\pt{X;\vec{k}_1,\vec{k_2}\vert \vec{x}_1,\vec{x_2}}=\frac{1}{2}\left\vert \psi\pt{\vec{k}_1} \psi\pt{\vec{k}_2} \right\vert^2\pq{1+\alpha\pt{X}\nu\cos\pt{\Delta\vec{x}\cdot\Delta\vec{k}}},
\end{equation}
where $\alpha\pt{A}=-1$ and $\alpha\pt{B}=1$. We observe quantum beats in $\Delta \vec{k}$ with a visibility that depends on $\nu$. In fact, for $\nu=0$, i.e. for two orthogonal photon states, we cannot observe quantum beats, since there are no interference effects. The visibility of the quantum beats is not related to the parameters $\vec{K}=\frac{\vec{k}_1+\vec{k}_2}{2}$. Therefore, detecting only $\Delta \vec{k}$ is necessary for the estimation of $\Delta \vec{x}$. The probability of having bunching or anti-bunching, while measuring $\pt{\vec{k}_1,\vec{k}_2}$ is
\begin{equation}
   P\pt{X;\vec{k}_1,\vec{k_2}\vert \vec{x}_1,\vec{x_2}}=\frac{1}{2}C\pt{\Delta \vec{k}}\pq{1+\alpha\pt{X}\nu\cos\pt{\Delta\vec{x}\cdot\Delta\vec{k}}},
\end{equation}
where the beats envelope $C\pt{\Delta \vec{k}}$ is
\begin{equation}
    C\pt{\Delta \vec{k}}=\int d^3K \left\vert \psi\pt{\vec{K}+\frac{\Delta \vec{k}}{2}} \psi\pt{\vec{K}-\frac{\Delta \vec{k}}{2}} \right\vert^2.\label{app:C}
\end{equation}
Therefore, if we consider a Gaussian distribution in the momentum space centered in $\vec{k}_0$ and with variance $\sigma^2_\omega,\sigma^2_{k_x},\sigma^2_{k_y}$ for each component of the vector $\vec{k}=(\omega,k_x,k_y)$,
\begin{equation}
    \left\vert \psi\pt{\vec{k}}\right\vert^2=\frac{1}{(2\pi)^{3/2}\sigma_\omega\sigma_{k_x}\sigma_{k_y}}\exp\left[-\frac{(\omega-\omega_0)^2}{2\sigma_\omega^2}\right]\exp\left[-\frac{(k_x-k_{x,0})^2}{2\sigma_{k_x}^2}\right]\exp\left[-\frac{(k_y-k_{y,0})^2}{2\sigma_{k_y}^2}\right],
\end{equation}
the beats envelope is
\begin{equation}
    C\pt{\Delta \vec{k}}=\frac{1}{(4\pi)^{3/2}\sigma_\omega\sigma_{k_x}\sigma_{k_y}}\exp\left[-\frac{\Delta\omega^2}{4\sigma_\omega^2}\right]\exp\left[-\frac{\Delta k_x^2}{4\sigma_{k_x}^2}\right]\exp\left[-\frac{\Delta k_y^2}{4\sigma_{k_y}^2}\right].
\end{equation}
By remembering that a photon with transverse momentum defined by $(k_x,k_y)$ is detected by the camera $C_j$ at $(x'_j,y'_j)$, such that $(k_x,k_y)= (x'_j,y'_j)\omega/cd$, the error in the detection for $(k_x,k_y)$, namely $\delta (k_x,k_y)$, is related to the errors of the position $\delta (x',y')$ and the frequency $\delta\omega$ as follows
\begin{equation}
    \delta k_x^2=k_x^2\pt{\frac{\delta x'^2}{x'^2_j}+\frac{\delta\omega^2}{\omega^2}},\qquad\delta k_y^2=k_y^2\pt{\frac{\delta y'^2}{y'^2_j}+\frac{\delta\omega^2}{\omega^2}}.
\end{equation}
The estimation of the parameters $\Delta\vec{x}$ is possible only under the condition that the errors in the measurement of the parameters are small enough to resolve the beating oscillations and to resolve both the transverse-momentum and frequency distributions of the two photons. These constraints are here summarized for the case of Gaussian probability distribution in frequency and transverse momentum:
\begin{equation}
        \delta \omega\ll \sigma_{\omega},\qquad \delta k_x\ll \sigma_{k_x},\qquad \delta k_y\ll \sigma_{k_y},\qquad \vert\Delta \vec{x}\delta \vec{k}\vert\ll 1.\label{app:condition}
\end{equation}

We consider now the presence of losses in the detection. We assume that both the efficiencies of the two detectors are equal to $\gamma\in\pq{0,1}$. Each photon can be lost in the detection with probability $1-\gamma$. Therefore, the probabilities of detecting zero, one, or two photons are, respectively, 
\begin{align}
    \begin{split}
       & P_0=\pt{1-\gamma}^2,\\
       & P_1=2\gamma\pt{1-\gamma},\\
       & P\pt{X;\Delta \vec{k}\vert \Delta \vec{x}}=\frac{\gamma^2}{2}C\pt{\Delta \vec{k}}\pq{1+\alpha\pt{X}\nu\cos\pt{\Delta \vec{x}\cdot\Delta \vec{k}}}.\label{app:resprob}
    \end{split}
\end{align}
The zero-photon and one-photon events do not carry any information for the estimation of $\Delta \vec{x}$. In fact, only the two-photon events have a probability that is function of these two parameters. This probability is proportional to $\gamma^2$. In particular, for $\nu=1$, the probabilities of detecting both photons in a bunching or an anti-bunching event are
\begin{align}
    \begin{split}
        P_{\nu=1}\pt{A;\Delta \vec{k}\vert \Delta \vec{x}}&=\gamma^2C\pt{\vec{k}}\sin^2\pt{\frac{\Delta \vec{x}\cdot\Delta \vec{k}}{2}},\\
        P_{\nu=1}\pt{B;\Delta \vec{k}\vert \Delta \vec{x}}&=\gamma^2C\pt{\vec{k}}\cos^2\pt{\frac{\Delta \vec{x}\cdot\Delta \vec{k}}{2}}.
    \end{split}
\end{align}
We define 
\begin{align}
\begin{split}
    \xi=\frac{\Delta\omega}{2\sigma_\omega}&, \qquad\tau=2\sigma_\omega\Delta t,\\
    \kappa_x=\frac{\Delta k_x}{2\sigma_{k_x}}&, \qquad\lambda_x=2\sigma_{k_x}\Delta x,\\
    \kappa_y=\frac{\Delta k_y}{2\sigma_{k_y}}&,\qquad\lambda_y=2\sigma_{k_y}\Delta y,
    \end{split}
\end{align}
recalled in the main text with Eq.~\ref{eq:notation}. Also, we introduce $\vec{\kappa}=(\xi,\kappa_x,\kappa_y)$,
\begin{equation}
s=\sqrt{\tau^2+\lambda_x^2+\lambda_y^2}=2\sqrt{\sigma_\omega^2\Delta t^2+\sigma_{k_x}^2\Delta x^2+\sigma_{k_y}^2\Delta y^2},
\end{equation}
and the unitary vector
\begin{equation}
   \vec{\lambda}= \frac{1}{\sqrt{\tau^2+\lambda_x^2+\lambda_y^2}}\begin{pmatrix}
        \tau\\-\lambda_x\\-\lambda_y
    \end{pmatrix}=\begin{pmatrix}
        \cos(\theta)\\\sin(\theta)\cos(\phi)\\\sin(\theta\sin(\phi))
    \end{pmatrix},
\end{equation}
that in the main text is introduced in Eq.~\ref{eq:unitaryvec}. Here, $\theta\in[0,\pi/2]$ and $\phi\in[0,2\pi[$, since we are fixing the sign of $\tau$ as positive. This choice is made because it is not possible to determine all the signs of $\tau,\lambda_x,\lambda_y$ simultaneously, because both $\vec{\lambda}$ and $-\vec{\lambda}$ give the same output probabilities. The parameters $\theta$ is
\begin{align}
    \theta=
        \arctan\frac{\sqrt{\lambda_x^2+\lambda_y^2}}{\tau}=\arctan\frac{\sqrt{\sigma_{k_x}^2\Delta x^2+\sigma_{k_y}^2\Delta y^2}}{\sigma_\omega \Delta t},
        \end{align}
        while $\phi$ is 
        \begin{align}
    &\phi=\arctan\frac{\lambda_y}{\lambda_x}=\arctan\frac{\sigma_{k_y}\Delta y}{\sigma_{k_x}\Delta x},
\end{align}
where the signs of $\lambda_x$ and $\lambda_y$ are appropriately considered in order to define in which quadrant $\phi$ is located.

By using these new definitions, we can write the output probabilities as
\begin{equation}
     P_\nu\pt{X;\vec{\kappa}\vert \vec{\lambda}}=\frac{\gamma^2}{(\pi)^{3/2}}\mathrm{e}^{-\vert\vec{\kappa}\vert^2}\zeta_{X;\nu}(s\vec{\kappa}\cdot\vec{\lambda}),\label{app:newprob}
\end{equation}
which is Eq.~\ref{eq:prob} in the main text, where 
\begin{equation}
    \zeta_{X;\nu}(s\vec{\kappa}\cdot\vec{\lambda})=\frac{1}{2}\pt{1+\alpha\pt{X}\nu\cos\pt{s\vec{\kappa}\cdot\vec{\lambda}}},\label{app:quantumbeats}
\end{equation}
which is Eq.~\ref{eq:quantumbeats} in the main text.  
 Instead, the conditions in Eq.~\ref{app:condition} can be written in the form
\begin{equation}
        \delta \vec{\kappa}\ll 1,\qquad\delta \vec{\kappa}\cdot\vec{\lambda}\ll \frac{1}{s}.
\end{equation}
These conditions are reported as Eq.~\ref{eq:conditions} in the main text.

By using Eq.~\ref{app:newprob}, we can find the output probabilities for the non-resolving scheme by integrating with respect of the three components of $\vec{\kappa}$, obtaining
\begin{align}
    \begin{split}
      \mathcal{P}_\nu\pt{X\vert s}=\int d^3\kappa P_\nu\pt{X;\vec{\kappa}\vert \vec{\lambda}}=\frac{\gamma^2}{2}\left(1+\alpha(X)\nu\exp\left[-s^2/4\right]\right),\label{app:nonresprob}
    \end{split}
\end{align}
which is Eq.~\ref{eq:non-resprob} in the main text.
\section{Quantum Fisher information in Eq.\ref{eq:Q}}
\label{app:QFI}
In this section we evaluate the quantum Fisher information matrix, which will assume the form
\begin{equation}
    Q(s,\theta,\phi)=\begin{pmatrix}
        Q_{ss}&Q_{s\theta}&Q_{s\phi}\\
        Q_{s\theta}&Q_{\theta\theta}&Q_{\theta\phi}\\
        Q_{s\phi}&Q_{\theta\phi}&Q_{\phi\phi}
    \end{pmatrix}\label{appeq:Q}
    ,
\end{equation}
where, the elements of the matrix are by definition
\begin{align}
    \begin{split}
Q_{\zeta\mu}=4\mathrm{Re}\pq{\Braket{\frac{\partial\Phi}{\partial\zeta}}{\frac{\partial\Phi}{\partial\mu}}-\Braket{\frac{\partial\Phi}{\partial\zeta}}{\Phi}\Braket{\Phi}{\frac{\partial\Phi}{\partial\mu}}},
    \end{split}\label{app:Qzm}
\end{align}
where $\zeta,\mu=s,\theta,\phi$.
For clarity, we recall the expression of the input state, which is
\begin{align}
    \begin{split}
        \ket{\Phi}&=\int d^3k_1 \mathrm{e}^{i\vec{x}_1\cdot\vec{k}_1}\left\vert\psi\pt{\vec{k}_1}\right\vert\hat{a}^\dagger_1\pt{\vec{k}_1}\ket{0}\otimes\int d^3k_2 \mathrm{e}^{i\vec{x}_2\cdot\vec{k}_2}\left\vert\psi\pt{\vec{k}_2}\right\vert\hat{d}^\dagger_2\pt{\vec{k}_2}\ket{0}\\
        &=\int d^3k_1 d^3k_2 \mathrm{e}^{i\vec{x}_1\cdot\vec{k}_1}\mathrm{e}^{i\vec{x}_2\cdot\vec{k}_2}\left\vert\psi\pt{\vec{k}_1}\psi\pt{\vec{k}_2}\right\vert\hat{a}^\dagger_1\pt{\vec{k}_1}\hat{d}^\dagger_2\pt{\vec{k}_2}\ket{0}.
    \end{split}
\end{align}
Here, only the phase depends on $\vec{x}_j$, $j=1,2$, therefore, only the phase will depend on $s,\theta,\phi$. We can manipulate the argument of the phase in order to explicit the dependence on these parameters
\begin{align}
    \begin{split}\vec{x}_1\cdot\vec{k_1}+\vec{x}_2\cdot\vec{k_2}&=\frac{1}{2}\Delta \vec{x}\cdot\Delta\vec{k}+\vec{x}\cdot\vec{K}\\
    &=\frac{s}{2}\vec{\kappa}\cdot\vec{\lambda}+\vec{x}\cdot\vec{K}.
    \end{split}
\end{align}
Therefore, we have
\begin{align}
    \begin{split}
    &\ket{\frac{\partial\Phi}{\partial s}}=\frac{i}{2}\int d^3k_1 d^3k_2 \vec{\kappa}\cdot\vec{\lambda}\mathrm{e}^{i\vec{x}_1\cdot\vec{k}_1}\mathrm{e}^{i\vec{x}_2\cdot\vec{k}_2}\left\vert\psi\pt{\vec{k}_1}\psi\pt{\vec{k}_2}\right\vert\hat{a}^\dagger_1\pt{\vec{k}_1}\hat{d}^\dagger_2\pt{\vec{k}_2}\ket{0},\\
        &\ket{\frac{\partial\Phi}{\partial \theta}}=\frac{is}{2}\int d^3k_1 d^3k_2 \vec{\kappa}\cdot\frac{\partial\vec{\lambda}}{\partial\theta}\mathrm{e}^{i\vec{x}_1\cdot\vec{k}_1}\mathrm{e}^{i\vec{x}_2\cdot\vec{k}_2}\left\vert\psi\pt{\vec{k}_1}\psi\pt{\vec{k}_2}\right\vert\hat{a}^\dagger_1\pt{\vec{k}_1}\hat{d}^\dagger_2\pt{\vec{k}_2}\ket{0},\\
        &\ket{\frac{\partial\Phi}{\partial \phi}}=\frac{is}{2}\int d^3k_1 d^3k_2 \vec{\kappa}\cdot\frac{\partial\vec{\lambda}}{\partial\phi}\mathrm{e}^{i\vec{x}_1\cdot\vec{k}_1}\mathrm{e}^{i\vec{x}_2\cdot\vec{k}_2}\left\vert\psi\pt{\vec{k}_1}\psi\pt{\vec{k}_2}\right\vert\hat{a}^\dagger_1\pt{\vec{k}_1}\hat{d}^\dagger_2\pt{\vec{k}_2}\ket{0}.
    \end{split}
\end{align}
Therefore, it is possible to evaluate the inner products that define the quantum Fisher information matrix. We obtain
\begin{align}
    \begin{split}
        \begin{pmatrix}
            \Braket{\frac{\partial\Phi}{\partial s}}{\frac{\partial\Phi}{\partial s}}&\Braket{\frac{\partial\Phi}{\partial s}}{\frac{\partial\Phi}{\partial \theta}}&\Braket{\frac{\partial\Phi}{\partial s}}{\frac{\partial\Phi}{\partial \phi}}\\
            \Braket{\frac{\partial\Phi}{\partial \theta}}{\frac{\partial\Phi}{\partial s}}&\Braket{\frac{\partial\Phi}{\partial \theta}}{\frac{\partial\Phi}{\partial \theta}}&\Braket{\frac{\partial\Phi}{\partial \theta}}{\frac{\partial\Phi}{\partial \phi}}\\
            \Braket{\frac{\partial\Phi}{\partial \phi}}{\frac{\partial\Phi}{\partial s}}&\Braket{\frac{\partial\Phi}{\partial \phi}}{\frac{\partial\Phi}{\partial \theta}}&\Braket{\frac{\partial\Phi}{\partial \phi}}{\frac{\partial\Phi}{\partial \phi}}
        \end{pmatrix}&=\frac{1}{4}\int d^3k_1 d^3k_2\left\vert\psi\pt{\vec{k}_1}\psi\pt{\vec{k}_2}\right\vert^2\times\\&\times\begin{pmatrix}
       (\vec{\kappa}\cdot\vec{\lambda})^2&s(\vec{\kappa}\cdot\vec{\lambda}) \left(\vec{\kappa}\cdot\frac{\partial\vec\lambda}{\partial\theta}\right)&s(\vec{\kappa}\cdot\vec{\lambda}) \left(\vec{\kappa}\cdot\frac{\partial\vec\lambda}{\partial\phi}\right)\\
       s(\vec{\kappa}\cdot\vec{\lambda}) \left(\vec{\kappa}\cdot\frac{\partial\vec\lambda}{\partial\theta}\right)&s^2\left(\vec{\kappa}\cdot\frac{\partial\vec\lambda}{\partial\theta}\right)^2&s^2\left(\vec{\kappa}\cdot\frac{\partial\vec\lambda}{\partial\theta}\right)\left(\vec{\kappa}\cdot\frac{\partial\vec\lambda}{\partial\phi}\right)\\
       s(\vec{\kappa}\cdot\vec{\lambda}) \left(\vec{\kappa}\cdot\frac{\partial\vec\lambda}{\partial\phi}\right)&s^2\left(\vec{\kappa}\cdot\frac{\partial\vec\lambda}{\partial\theta}\right)\left(\vec{\kappa}\cdot\frac{\partial\vec\lambda}{\partial\phi}\right)&s^2\left(\vec{\kappa}\cdot\frac{\partial\vec\lambda}{\partial\phi}\right)^2
    \end{pmatrix}.
    \end{split}
\end{align}
By using Eq.~\ref{app:C}, we obtain
\begin{align}
    \begin{split}
        \begin{pmatrix}
            \Braket{\frac{\partial\Phi}{\partial s}}{\frac{\partial\Phi}{\partial s}}&\Braket{\frac{\partial\Phi}{\partial s}}{\frac{\partial\Phi}{\partial \theta}}&\Braket{\frac{\partial\Phi}{\partial s}}{\frac{\partial\Phi}{\partial \phi}}\\
            \Braket{\frac{\partial\Phi}{\partial \theta}}{\frac{\partial\Phi}{\partial s}}&\Braket{\frac{\partial\Phi}{\partial \theta}}{\frac{\partial\Phi}{\partial \theta}}&\Braket{\frac{\partial\Phi}{\partial \theta}}{\frac{\partial\Phi}{\partial \phi}}\\
            \Braket{\frac{\partial\Phi}{\partial \phi}}{\frac{\partial\Phi}{\partial s}}&\Braket{\frac{\partial\Phi}{\partial \phi}}{\frac{\partial\Phi}{\partial \theta}}&\Braket{\frac{\partial\Phi}{\partial \phi}}{\frac{\partial\Phi}{\partial \phi}}
        \end{pmatrix}&=\frac{1}{4(\pi)^{3/2}}\int d^3\kappa\mathrm{e^{-\vert\kappa\vert^2}}\begin{pmatrix}
       (\vec{\kappa}\cdot\vec{\lambda})^2&s(\vec{\kappa}\cdot\vec{\lambda}) \left(\vec{\kappa}\cdot\frac{\partial\vec\lambda}{\partial\theta}\right)&s(\vec{\kappa}\cdot\vec{\lambda}) \left(\vec{\kappa}\cdot\frac{\partial\vec\lambda}{\partial\phi}\right)\\
       s(\vec{\kappa}\cdot\vec{\lambda}) \left(\vec{\kappa}\cdot\frac{\partial\vec\lambda}{\partial\theta}\right)&s^2\left(\vec{\kappa}\cdot\frac{\partial\vec\lambda}{\partial\theta}\right)^2&s^2\left(\vec{\kappa}\cdot\frac{\partial\vec\lambda}{\partial\theta}\right)\left(\vec{\kappa}\cdot\frac{\partial\vec\lambda}{\partial\phi}\right)\\
       s(\vec{\kappa}\cdot\vec{\lambda}) \left(\vec{\kappa}\cdot\frac{\partial\vec\lambda}{\partial\phi}\right)&s^2\left(\vec{\kappa}\cdot\frac{\partial\vec\lambda}{\partial\theta}\right)\left(\vec{\kappa}\cdot\frac{\partial\vec\lambda}{\partial\phi}\right)&s^2\left(\vec{\kappa}\cdot\frac{\partial\vec\lambda}{\partial\phi}\right)^2
    \end{pmatrix}.
    \end{split}\label{app:QFIpart1}
\end{align}
Starting from the term $\Braket{\frac{\partial\Phi}{\partial s}}{\frac{\partial\Phi}{\partial \theta}}$, we see that $\vec{\lambda}\cdot\partial\vec{\lambda}/\partial\theta=0$ and $\vert\vec{\lambda}\vert=\vert\partial\vec{\lambda}/\partial\theta\vert=1$. Therefore, we can rotate the integration parameters in such a way $\vec{\kappa}\cdot\vec{\lambda}=\kappa_x$ and $\vec{\kappa}\cdot\partial\vec{\lambda}/\partial\theta=\kappa_y$. This means that
\begin{equation}
    \Braket{\frac{\partial\Phi}{\partial s}}{\frac{\partial\Phi}{\partial \theta}}=\frac{s}{4(\pi)^{3/2}}\int d^3\kappa\mathrm{e}^{-\vert\vec{\kappa}\vert^2}\kappa_x \kappa_y.
\end{equation}
Since the integrand is an odd function with respect to the integration parameters $\kappa_x$ and $\kappa_y$, and the integration domain is $\mathbb{R}^2$, we have $\Braket{\frac{\partial\Phi}{\partial s}}{\frac{\partial\Phi}{\partial \theta}}=0$. A similar approach can be used to prove that $\Braket{\frac{\partial\Phi}{\partial s}}{\frac{\partial\Phi}{\partial \phi}}=\Braket{\frac{\partial\Phi}{\partial \phi}}{\frac{\partial\Phi}{\partial \theta}}=0$.
We can find the diagonal terms of the Fisher information matrix by rotating appropriately the integration parameters with the same way we used for the evaluation of the off-diagonal terms. We obtain
\begin{align}
\begin{split}
    \Braket{\frac{\partial\Phi}{\partial s}}{\frac{\partial\Phi}{\partial s}}&=\frac{1}{4(\pi)^{3/2}}\int d^3\kappa\mathrm{e}^{-\vert\vec{\kappa}\vert^2}
       (\vec{\kappa}\cdot\vec{\lambda})^2\\
       &=\frac{1}{4(\pi)^{3/2}}\int d^3\kappa\mathrm{e}^{-\vert\vec{\kappa}\vert^2}
       \kappa_x^2=\frac{1}{8}.
       \end{split}
\end{align}
\begin{align}
\begin{split}
    \Braket{\frac{\partial\Phi}{\partial \theta}}{\frac{\partial\Phi}{\partial \theta}}&=\frac{s^2}{4(\pi)^{3/2}}\int d^3\kappa\mathrm{e}^{-\vert\vec{\kappa}\vert^2}
       \left(\vec{\kappa}\cdot\frac{\partial\vec\lambda}{\partial\theta}\right)^2\\
       &=\frac{s^2}{4(\pi)^{3/2}}\int d^3\kappa\mathrm{e}^{-\vert\vec{\kappa}\vert^2}
       \kappa_y^2=\frac{s^2}{8}.
       \end{split}
\end{align}
\begin{align}
\begin{split}
    \Braket{\frac{\partial\Phi}{\partial \phi}}{\frac{\partial\Phi}{\partial \phi}}&=\frac{s^2}{4(\pi)^{3/2}}\int d^3\kappa\mathrm{e}^{-\vert\vec{\kappa}\vert^2}
       \left(\vec{\kappa}\cdot\frac{\partial\vec\lambda}{\partial\phi}\right)^2\\
       &=\frac{s^2\sin^2\theta}{4(\pi)^{3/2}}\int d^3\kappa\mathrm{e}^{-\vert\vec{\kappa}\vert^2}
       \kappa_y^2=\frac{s^2\sin^2\theta}{8}.
       \end{split}
\end{align}
Therefore Eq.\ref{app:QFIpart1} became
\begin{equation}
    \begin{pmatrix}
            \Braket{\frac{\partial\Phi}{\partial s}}{\frac{\partial\Phi}{\partial s}}&\Braket{\frac{\partial\Phi}{\partial s}}{\frac{\partial\Phi}{\partial \theta}}&\Braket{\frac{\partial\Phi}{\partial s}}{\frac{\partial\Phi}{\partial \phi}}\\
            \Braket{\frac{\partial\Phi}{\partial \theta}}{\frac{\partial\Phi}{\partial s}}&\Braket{\frac{\partial\Phi}{\partial \theta}}{\frac{\partial\Phi}{\partial \theta}}&\Braket{\frac{\partial\Phi}{\partial \theta}}{\frac{\partial\Phi}{\partial \phi}}\\
            \Braket{\frac{\partial\Phi}{\partial \phi}}{\frac{\partial\Phi}{\partial s}}&\Braket{\frac{\partial\Phi}{\partial \phi}}{\frac{\partial\Phi}{\partial \theta}}&\Braket{\frac{\partial\Phi}{\partial \phi}}{\frac{\partial\Phi}{\partial \phi}}
        \end{pmatrix}=\frac{1}{8}\begin{pmatrix}
            1&0&0\\0&s^2&0\\0&0&s^2\sin^2\theta
        \end{pmatrix}.\label{app:QFIquarti}
\end{equation}
For the evaluation of the quantum Fisher information in Eq.\ref{app:Qzm}, we evaluate also
\begin{align}
    \begin{split}
      \Braket{\Phi}{\frac{\partial\Phi}{\partial s}}&=\frac{i}{2}\int d^3k_1 d^3k_2 \vec{\kappa}\cdot\vec{\lambda}\left\vert\psi\pt{\vec{k}_1}\psi\pt{\vec{k}_2}\right\vert^2,\\\Braket{\Phi}{\frac{\partial\Phi}{\partial \theta}}&=\frac{is}{2}\int d^3k_1 d^3k_2 \vec{\kappa}\cdot\frac{\partial\vec{\lambda}}{\partial\theta}\left\vert\psi\pt{\vec{k}_1}\psi\pt{\vec{k}_2}\right\vert^2,\\\Braket{\Phi}{\frac{\partial\Phi}{\partial \phi}}&=\frac{is}{2}\int d^3k_1 d^3k_2 \vec{\kappa}\cdot\frac{\partial\vec{\lambda}}{\partial\phi}\left\vert\psi\pt{\vec{k}_1}\psi\pt{\vec{k}_2}\right\vert^2.
    \end{split}
\end{align}
By using Eq.~\ref{app:C} we obtain
\begin{align}
    \begin{split}
      \Braket{\Phi}{\frac{\partial\Phi}{\partial s}}&=\frac{i}{2(\pi)^{3/2}}\int d^3\kappa\mathrm{e^{-\vert\kappa\vert^2}} \vec{\kappa}\cdot\vec{\lambda}=0,\\\Braket{\Phi}{\frac{\partial\Phi}{\partial \theta}}&=\frac{is}{2(\pi)^{3/2}}\int d^3\kappa\mathrm{e^{-\vert\kappa\vert^2}} \vec{\kappa}\cdot\frac{\partial\vec{\lambda}}{\partial\theta}=0,\\\Braket{\Phi}{\frac{\partial\Phi}{\partial \phi}}&=\frac{is}{2(\pi)^{3/2}}\int d^3\kappa\mathrm{e^{-\vert\kappa\vert^2}} \vec{\kappa}\cdot\frac{\partial\vec{\lambda}}{\partial\phi}=0.
    \end{split}
\end{align}
Therefore the only non-zero term of the Quantum Fisher information in Eq.~\ref{app:Qzm} is given by Eq.~\ref{app:QFIquarti}. Its result is
\begin{equation}
    Q(s,\theta,\phi)=\frac{1}{2}\begin{pmatrix}
        1&0&0\\0&s^2&0\\0&0&s^2\sin^2\theta
    \end{pmatrix},
\end{equation}
which is the Eq.~\ref{eq:Q} in the main text.
\section{Evaluation of Eqs.~\ref{eq:finu1},~\ref{eq:finu1s0},\ref{eq:Fi},~\ref{eq:densityfi},~\ref{eq:beta}}
\label{app:FI}
In this section we will analyze in detail the Fisher information matrix for the sensing scheme, its density, and the Fisher information matrix for the non-resolving protocol. 

By using Eq.~\ref{app:newprob}, the Fisher information matrix for the parameters $s,\theta,\phi$ is the following
\begin{align}
\begin{split}
    F_\nu(s,\theta,\phi)&=\begin{pmatrix}
        F_{ss}&F_{s\theta}&F_{s\phi}\\
        F_{s\theta}&F_{\theta\theta}&F_{\theta\phi}\\
        F_{s\phi}&F_{\theta\phi}&F_{\phi\phi}
    \end{pmatrix}\\&=\frac{\gamma^2}{(\pi)^{3/2}}\int d^3\kappa\mathrm{e}^{-\vert\vec{\kappa}\vert^2}\beta(s\vec{\kappa}\cdot\vec{\lambda})\begin{pmatrix}
       (\vec{\kappa}\cdot\vec{\lambda})^2&s(\vec{\kappa}\cdot\vec{\lambda}) \left(\vec{\kappa}\cdot\frac{\partial\vec\lambda}{\partial\theta}\right)&s(\vec{\kappa}\cdot\vec{\lambda}) \left(\vec{\kappa}\cdot\frac{\partial\vec\lambda}{\partial\phi}\right)\\
       s(\vec{\kappa}\cdot\vec{\lambda}) \left(\vec{\kappa}\cdot\frac{\partial\vec\lambda}{\partial\theta}\right)&s^2\left(\vec{\kappa}\cdot\frac{\partial\vec\lambda}{\partial\theta}\right)^2&s^2\left(\vec{\kappa}\cdot\frac{\partial\vec\lambda}{\partial\theta}\right)\left(\vec{\kappa}\cdot\frac{\partial\vec\lambda}{\partial\phi}\right)\\
       s(\vec{\kappa}\cdot\vec{\lambda}) \left(\vec{\kappa}\cdot\frac{\partial\vec\lambda}{\partial\phi}\right)&s^2\left(\vec{\kappa}\cdot\frac{\partial\vec\lambda}{\partial\theta}\right)\left(\vec{\kappa}\cdot\frac{\partial\vec\lambda}{\partial\phi}\right)&s^2\left(\vec{\kappa}\cdot\frac{\partial\vec\lambda}{\partial\phi}\right)^2
    \end{pmatrix},\label{app:fidef}
    \end{split}
\end{align}
where, $\beta(s\vec{\kappa}\cdot\vec{\lambda})$ is obtained by using Eq.~\ref{app:quantumbeats},
\begin{equation}
    \beta_\nu(x)=\sum_{X=A,B}\frac{1}{\zeta_{X;\nu}(x)}\left(\frac{\partial \zeta_{X;\nu}(x)}{\partial x}\right)^2=\frac{\nu^2\sin^2(x)}{1-\nu^2\cos^2(x)},
\end{equation}
which is Eq.~\ref{eq:beta} of the main text.

Starting from the term $F_{s\theta}$, we see that $\vec{\lambda}\cdot\partial\vec{\lambda}/\partial\theta=0$ and $\vert\vec{\lambda}\vert=\vert\partial\vec{\lambda}/\partial\theta\vert=1$. Therefore, we can rotate the integration parameters in such a way $\vec{\kappa}\cdot\vec{\lambda}=\kappa_x$ and $\vec{\kappa}\cdot\partial\vec{\lambda}/\partial\theta=\kappa_y$. This means that
\begin{equation}
    F_{s\theta}=\frac{\gamma^2s}{(\pi)^{3/2}}\int d^3\kappa\mathrm{e}^{-\vert\vec{\kappa}\vert^2}\beta(s\kappa_x)\kappa_x \kappa_y.
\end{equation}
Since the integrand is an odd function with respect to the integration parameters $k_x$ and $k_y$, and the integration domain is $\mathbb{R}^2$, we have $F_{s\theta}=0$. A similar approach can be used to prove that $F_{s\phi}=F_{\theta\phi}=0$.

We can find the diagonal terms of the Fisher information matrix by rotating appropriately the integration parameters with the same way we used for the evaluation of the off-diagonal terms. We obtain
\begin{align}
\begin{split}
    F_{ss}&=\frac{\gamma^2}{(\pi)^{3/2}}\int d^3\kappa\mathrm{e}^{-\vert\vec{\kappa}\vert^2}\beta(s\vec{\kappa}\cdot\vec{\lambda})
       (\vec{\kappa}\cdot\vec{\lambda})^2\\
       &=\frac{\gamma^2}{(\pi)^{3/2}}\int d^3\kappa\mathrm{e}^{-\vert\vec{\kappa}\vert^2}\beta(s\kappa_x)
       \kappa_x^2\\&
       =\frac{\gamma^2}{(\pi)^{1/2}}\int d\kappa_x\mathrm{e}^{-\kappa_x^2}\beta(s\kappa_x)
       \kappa_x^2.
       \end{split}
\end{align}
\begin{align}
\begin{split}
    F_{\theta\theta}&=\frac{\gamma^2s^2}{(\pi)^{3/2}}\int d^3\kappa\mathrm{e}^{-\vert\vec{\kappa}\vert^2}\beta(s\vec{\kappa}\cdot\vec{\lambda})
       \left(\vec{\kappa}\cdot\frac{\partial\vec\lambda}{\partial\theta}\right)^2\\
       &=\frac{\gamma^2s^2}{(\pi)^{3/2}}\int d^3\kappa\mathrm{e}^{-\vert\vec{\kappa}\vert^2}\beta(s\kappa_x)
       \kappa_y^2\\&
       =\frac{\gamma^2s^2}{2(\pi)^{1/2}}\int d\kappa_x\mathrm{e}^{-\kappa_x^2}\beta(s\kappa_x).
       \end{split}
\end{align}
\begin{align}
\begin{split}
    F_{\phi\phi}&=\frac{\gamma^2s^2}{(\pi)^{3/2}}\int d^3\kappa\mathrm{e}^{-\vert\vec{\kappa}\vert^2}\beta(s\vec{\kappa}\cdot\vec{\lambda})
       \left(\vec{\kappa}\cdot\frac{\partial\vec\lambda}{\partial\phi}\right)^2\\
       &=\frac{\gamma^2s^2\sin^2\theta}{(\pi)^{3/2}}\int d^3\kappa\mathrm{e}^{-\vert\vec{\kappa}\vert^2}\beta(s\kappa_x)
       \kappa_y^2\\&
       =\frac{\gamma^2s^2\sin^2\theta}{2(\pi)^{1/2}}\int d\kappa_x\mathrm{e}^{-\kappa_x^2}\beta(s\kappa_x).
       \end{split}
\end{align}
As a result, the Fisher information matrix in Eq.~\ref{app:fidef} is
\begin{align}
F_{\nu}(s,\theta,\phi)&=\gamma^2\int d\rho f_\nu(\rho;s,\theta,\phi),
\end{align}
where
\begin{equation}
    f_\nu(\rho;s,\theta,\phi)= \frac{\mathrm{e}^{-\rho^2}\beta_\nu(s\rho)}{(\pi)^{1/2}}\begin{pmatrix}
        \rho^2&0&0\\
        0&\frac{s^2}{2}&0\\
        0&0&\frac{s^2\sin^2\theta}{2}
    \end{pmatrix}.
\end{equation}
These last two equations are Eq.~\ref{eq:Fi} and Eq.~\ref{eq:densityfi} in the main text. For $\nu=1$, we obtain Eq.~\ref{eq:finu1} of the main text,
\begin{equation}
  F_{\nu=1}(s,\theta,\phi)=\frac{\gamma^2}{2}\begin{pmatrix}
        1&0&0\\
        0&s^2&0\\
        0&0&s^2\sin^2\theta
    \end{pmatrix}=\gamma^2Q(s,\theta,\phi) .
\end{equation}

Instead, the Fisher information for evaluation of $s$ by using the non-resolving scheme can be found by using Eq.~\ref{app:nonresprob}
\begin{equation}
    \mathcal{F}_\nu(s)=\frac{\gamma^2\nu^2}{\mathrm{e}^{s^2/2}-\nu^2}\frac{s^2}{4}.
\end{equation}
For maximum visibility, i.e. $\nu=1$, and $s\ll 1$, the exponential can be approximated as $\mathrm{e}^{s^2/2}\xrightarrow{s\rightarrow 0}1+s^2/2$, giving Eq.~\ref{eq:finu1s0} of the main text
\begin{equation}
    \mathcal{F}_{\nu=1}(s)\xrightarrow{s\rightarrow 0}\gamma^2Q_{ss}(s,\theta,\phi)=\frac{\gamma^2}{2}.
\end{equation}
This equation proves that for the estimation of $s$, when the visibility is maximum, the resolving protocol and the non-resolving protocol are equivalent.

\section{Likelihood estimation for the realization of FIG~\ref{fig:sim}}
\label{app:Likelihood}
In this section, a derivation of the likelihood estimators for the estimation of the parameters $(s,\theta,\phi)$ follows. The sensing scheme presented in this work is based on resolved-sampling measures. By considering that the i-th element of the sample has a probability $P_\nu\pt{X_i;\vec{\kappa}_i\vert s,\vec{\lambda}}  
$ to occurs, we define the likelihood as
\begin{equation}
\mathcal{L}=\prod_i P_\nu\pt{X_i;\vec{\kappa}_i\vert s,\vec{\lambda}},  
\end{equation}
and the Log-Likelihood, as it logarithm, i.e.
\begin{equation}
\log\mathcal{L}\propto\sum_i\log\pq{1+\alpha\pt{X_i}\nu\cos\pt{s\vec{\kappa}_i\cdot\vec{\lambda}}}.
\end{equation}
In order to estimate the parameters $(s,\theta,\phi)$, we define three estimators such that the Log-Likelihood, as a function of these estimators, is maximized. Therefore, we impose the stationariety condition, for which
\begin{align}
    \begin{split}
        \begin{cases}
            \displaystyle\frac{\partial\log\mathcal{L}}{\partial s}=0\\
            \\\displaystyle\frac{\partial\log\mathcal{L}}{\partial \theta}=0\\
            \\\displaystyle\frac{\partial\log\mathcal{L}}{\partial \phi}=0
        \end{cases}\Longrightarrow
                \begin{cases}
           \displaystyle\sum_i \frac{\alpha\pt{X_i}\nu\sin\pt{s\vec{\kappa}_i\cdot\vec{\lambda}}}{1+\alpha\pt{X_i}\nu\cos\pt{s\vec{\kappa}_i\cdot\vec{\lambda}}}\vec{\kappa}_i\cdot\vec{\lambda}=0\\
           \\
                      \displaystyle\sum_i \frac{\alpha\pt{X_i}\nu\sin\pt{s\vec{\kappa}_i\cdot\vec{\lambda}}}{1+\alpha\pt{X_i}\nu\cos\pt{s\vec{\kappa}_i\cdot\vec{\lambda}}}\vec{\kappa}_i\cdot\frac{\partial\vec{\lambda}}{\partial\theta}=0\\
                      \\
                      \displaystyle\sum_i \frac{\alpha\pt{X_i}\nu\sin\pt{s\vec{\kappa}_i\cdot\vec{\lambda}}}{1+\alpha\pt{X_i}\nu\cos\pt{s\vec{\kappa}_i\cdot\vec{\lambda}}}\vec{\kappa}_i\cdot\frac{\partial\vec{\lambda}}{\partial\phi}=0
        \end{cases}.
    \end{split}
\end{align}
Since $\vec{\lambda}$, $\frac{\partial\vec{\lambda}}{\partial\theta}$ and $\frac{\partial\vec{\lambda}}{\partial\phi}$ are three orthogonal vectors, solving this system is equivalent to solve
\begin{equation}
  \displaystyle\sum_i \frac{\alpha\pt{X_i}\nu\sin\pt{s\vec{\kappa}_i\cdot\vec{\lambda}}}{1+\alpha\pt{X_i}\nu\cos\pt{s\vec{\kappa}_i\cdot\vec{\lambda}}}\vec{\kappa}_i=\begin{pmatrix}
      0\\0\\0
  \end{pmatrix}  .
\end{equation}
By numerically solving this system for $s,\theta,\phi$ as functions of $\{ X_i,\vec{\kappa}_i\}$, it is possible to find the likelihood estimators. 

\end{document}